\input harvmac
\def\NSVZ{{\rm NSVZ}}
\def\DRED{{\rm DRED}}
\def\npb{{Nucl.\ Phys.\ }{\bf B}}
 
\def\prd{{Phys.\ Rev.\ }{\bf D}}
\def\prl{Phys.\ Rev.\ Lett.\ }
\def\plb{{Phys.\ Lett.\ }{\bf B}}
\def\ijmpa{{Int.\ J.\ Mod.\ Phys.\ }{\bf A}}
\def\Ph{\Phi}

\def\th{\theta}
\def\bM{M^*}

\def\bPh{\bar\Ph}
\def\be{\bar\eta}

\def\tY{\tilde Y}

\def\bt{\tilde b}
\def\lf{16\pi^2}
\def\llf{(16\pi^2)^2}
\def\lllf{(16\pi^2)^3}
{\nopagenumbers
\line{\hfil LTH 408}
\line{\hfil hep-ph/9709364}
\vskip .5in
\centerline{\titlefont The Gaugino $\beta$-Function}
\vskip 1in
\centerline{\bf I.~Jack and D.R.T.~Jones }
\medskip   
\centerline{\it Dept. of Mathematical Sciences,
University of Liverpool, Liverpool L69 3BX, UK}
\vskip .3in
We present an elegant exact formula for the gaugino $\beta$-function
in a softly-broken supersymmetric gauge theory, of the form 
$\beta_M={\cal O}(\beta_g/g)$, where $\beta_g$ is the gauge $\beta$ function 
and ${\cal O}$ is a simple differential operator acting on the gauge coupling
$g$ and the Yukawa coupling. 
 
\Date{ September 1997}

The all-orders gauge $\beta$-function for a 
general  $N=1$ supersymmetric gauge theory
has been known 
for some time\ref\NSVZb{V.~Novikov et al, \npb 229 (1983) 381\semi
V.~Novikov et al, \plb166 (1986) 329\semi
M.~Shifman and A.~Vainstein, \npb 277 (1986) 456}. The early derivations
were based on anomaly arguments and instanton calculus. 
Later arguments emphasised the importance of 
holomorphy, and these ideas
have been further refined and explained under the impetus of 
developments in supersymmetric duality
\ref\Shif{M.~Shifman, \ijmpa11 (1996) 5761\semi
N.~Arkani-Hamed  and H.~Murayama, hep-th/9707133}. 
Recently similar ideas have been 
applied to softly-broken superymmetric gauge theories, and 
renormalisation-group invariant quantities involving the gaugino mass
were constructed\ref\HS{J.~Hisano and M.~Shifman, hep-ph/9705417}. 
We shall use these results to derive simple and  elegant 
expressions for the gaugino mass $\beta$-function, $\beta_M$. 
One of these (Eq.~(15))
is very analogous to the standard NSVZ result for $\beta_g$; the other 
(Eq.~(17)) expresses $\beta_M$ as a simple operator acting on $\beta_g$.
In fact, the action of this operator is equivalent to the application 
of a set of rules devised by Yamada\ref\yam{Y.~Yamada, \prd50 (1994) 3537}\
for obtaining the $\beta$-functions
for the scalar soft-breaking couplings starting from the anomalous 
dimension for the chiral superfields. We start by reviewing Yamada's rules,
then go on to derive our exact results for $\beta_M$, and discuss 
their scheme dependence. We illustrate the results with explicit 
results up to three loops. Finally we show that in a one-loop finite theory,
$\beta_g$ and $\beta_M$ 
can be made to vanish to all orders by a suitable choice of
renormalisation scheme.

Yamada's rules are based on 
the spurion formalism\ref\spur{L.~Girardello and M.T.~Grisaru,
\npb194 (1982) 65\semi
J.A.~Helay\"el-Neto, \plb135 (1984) 78\semi
F.~Feruglio, J.A.~Helay\"el-Neto and F.~Legovini, \npb249 (1985) 533\semi
M.~Scholl, Z.Phys.~{\bf C}28 (1985) 545}, which enables one to write 
the softly broken $N=1$ theory in terms of superfields. 
The lagrangian for the theory can be written
\eqn\Aa{
L(\Ph,W)=L_{SUSY}+L_{SB}+L_{GF}+L_{FP} }
where $L_{SUSY}$ is the usual $N=1$ supersymmetric lagrangian, with a 
superpotential $W(\Phi)={1\over6}Y^{ijk}\Phi_i\Phi_j\Phi_k+
{1\over2}\mu^{ij}\Phi_i\Phi_j$.
We assume a simple gauge group with no gauge singlet fields.  
The soft breaking part $L_{SB}$ may be written 
\eqn\Aba{\eqalign{
L_{SB}(\Ph,W)&=-\left\{\int d^2\th\eta\left({1\over6}h^{ijk}\Ph_i\Ph_j\Ph_k
+{1\over2}b^{ij}\Ph_i\Ph_j+{1\over2}MW_A{}^{\alpha}W_{A\alpha}\right)
+{\rm h.c.}\right\}\cr
&\quad -\int d^4\th\be\eta\bPh^j(m^2)^i{}_j(e^{2gV})_i{}^k\Ph_k,\cr}}
where $\eta=\th^2$ is the spurion external field.
The gauge-fixing and Fadeev-Popov terms are contained in $L_{GF}$ and
$L_{FP}$ respectively.  
It is convenient to introduce a generalised form $\gamma_{\eta}$ 
of the anomalous 
dimension $\gamma$ of the chiral supermultiplet, given by: 
\eqn\Ah{
\gamma_{\eta}=\gamma+\gamma_1\eta+\gamma_1^{\dagger}\be+\gamma_2\be\eta.} 
It was shown by Yamada\yam\ that $(\gamma_{\eta})^i{}_j$ could be obtained
from $(\gamma)^i{}_j$ by the following rules:
\item{1.} Replace $Y^{lmn}$ by $Y^{lmn}-h^{lmn}\eta$.
\item{2.} Replace the gauge coupling $g^2$ by $g^2(1+M\eta+\bM\be+2|M|^2\be 
\eta)$.
\item{3.} Insert $\delta^{l'}{}_l+(m^2)^{l'}{}_l\be\eta$ between contracted 
indices $l'$ and $l$ in $Y$ and $Y^*$, respectively: $Y^{lmn}Y_{lm'n'}
\rightarrow Y^{lmn}Y_{lm'n'}+Y^{lmn}(m^2)^{l'}{}_lY_{l'm'n'}\be\eta$ (where,
here and subsequently, $Y_{lmn}=(Y^{lmn})^*$).
\item{4.} Replace a term $T^i{}_j$ in 
$\gamma^i{}_j$ with no Yukawa couplings by 
$T^i{}_j-(m^2)^i{}_kT^k{}_j\be\eta$.

\noindent$\gamma_1$ and $\gamma_2$ may then 
be obtained by extracting the coefficients of $\eta$
and $\be\eta$ respectively.
In the case of $\gamma_1$,  
the above rules can be subsumed by the simple relation 
\eqn\Aj{
(\gamma_1)^i{}_j={\cal O}\gamma^i{}_j,}
where
\eqn\Ajb{
{\cal O}=\left(Mg^2{\partial\over{\partial g^2}}-h^{lmn}{\partial
\over{\partial Y^{lmn}}}\right).}
It is  straightforward to show that 
\eqn\Ai{\eqalign{
\beta_h^{ijk}&=\gamma^i{}_lh^{ljk}+\gamma^j{}_lh^{ilk}
+\gamma^k{}_lh^{ijl}-2\gamma_1^i{}_lY^{ljk}
-2\gamma_1^j{}_lY^{ilk}-2\gamma_1^k{}_lY^{ijl}\cr
\beta_b^{ij}&=\gamma^i{}_lb^{lj}+\gamma^j{}_lb^{il}
-2\gamma_1^i{}_l\mu^{lj}-2\gamma_1^j{}_l\mu^{il}.\cr}}
These results are similar in form to the standard results for $\beta_Y$
and $\beta_{\mu}$ which follow from the non-renormalisation theorem, namely
\eqn\Aia{\eqalign{
\beta_Y^{ijk}=&\gamma^i{}_lY^{ljk}+\gamma^j{}_lY^{ilk}+\gamma^k{}_lY^{ijl},\cr
\beta_{\mu}^{ij}=&\gamma^i{}_l\mu^{lj}+\gamma^j{}_l\mu^{il}.\cr}}

It also appears from Eqs.~\Aba\ and \Ah\ that
\eqn\Aib{
(\beta_{m^2})^i{}_j={1\over2}\gamma^i{}_k(m^2)^k{}_j
+{1\over2}(m^2)^i{}_k\gamma^k{}_j+\gamma_2^i{}_j,}
which we may write using Yamada's rules as 
\eqn\Ajc{
(\beta_{m^2})^i{}_j=\left[2{\cal O}{\cal O}^* +2M\bM g^2{\partial
\over{\partial g^2}} +\tY_{lmn}{\partial\over{\partial Y_{lmn}}}
+\tY^{lmn}{\partial\over{\partial Y^{lmn}}}\right]\gamma^i{}_j,}
where
\eqn\Ajd{
\tY^{ijk}=(m^2)^i{}_lY^{ljk}+(m^2)^j{}_lY^{ilk}+(m^2)^k{}_lY^{ijl}.}
As was discussed in Refs.~\ref\jj{I.~Jack and
D.R.T.~Jones, \plb 333 (1994) 372.}, \ref\jjmvy{I.~Jack, D.R.T~Jones, 
S.P.~Martin, M.T.~Vaughn and Y.~Yamada, \prd50 (1994) R5481}, however, 
when using
standard regularisation by dimensional reduction (DRED), Eq.~\Aib\ 
acquires additional terms arising from the presence of $\epsilon$-scalars.
This leaves open the possibility 
that Eq.~\Aib\ is correct as it stands in some 
scheme which does not require the introduction of $\epsilon$-scalars; we shall
discuss this in more detail later. 

Recently Hisano and Shifman (HS)\HS\ have found an exact renormalisation group 
invariance involving the gaugino mass. To make contact with their 
results, suppose that we can write
\eqn\Aja{
b^{ij}={1\over2}(\bt^i{}_l\mu^{lj}+\bt^j{}_l\mu^{il}).}
In our notation, the HS result is that the combination 
\eqn\Ak{
{M\over{g^2}}\left(1-{2C(G)g^2\over{16\pi^2}}\right)-{1\over{16\pi^2r}}
\tr[\bt C(R)],}
is RG invariant.
Here
\eqn\Aaca{
C(G)\delta_{AB} = f_{ACD}f_{BCD}, \quad C(R)^i{}_j = (R_A R_A)^i{}_j,\quad 
r=\delta_{AA}.}
 Our purpose here is to obtain an elegant formula for the gaugino mass 
$\beta$-function, and explore some of its consequences.
Firstly, note that if we assume that $[\bt,\gamma]=0$ then 
it follows from Eqs.~\Ai\ and \Aia\ that
\eqn\Al{
(\beta_{\bt})^i{}_j=-4(\gamma_1)^i{}_j.}
It follows immediately by taking $\mu{d\over{d \mu}}$ of Eq.~\Ak\ that 
\eqn\Am{
\beta_M^{\NSVZ}=2{{M\beta_g -2g^3\tr[\gamma_1C(R)](16\pi^2 r)^{-1}}\over{
g[1-2g^2C(G)(16\pi^2)^{-1}]}}.}
To derive this result, we need
the NSVZ result for $\beta_g$\NSVZb,
\eqn\An{
\beta_g^{\NSVZ}={g^3\over{16\pi^2}}\left[{Q-2r^{-1}\tr[\gamma C(R)]\over{
1-2g^2C(G)(\lf)^{-1}}}\right] ,}
where $Q=T(R)-3C(G)$ (with $T(R)$ defined by $T(R)\delta_{AB} = \Tr(R_A R_B)$).
The formal resemblance between Eq.~\An\  and Eq.~\Am\ is quite striking.  
The HS result Eq.~\Ak~\ appears to require the existence 
of a tensor $b^{ij}$, and in order to obtain Eq.~\Am\ we assumed that 
$[\bt,\gamma]=0$; $b$ does not appear in our result, however, 
and we claim it to be quite general. 
It is easy to show that $\beta_M$
in Eq.~\Am\ may be written in the following very compact form,
\eqn\Ao{
\beta_M=2{\cal O}\left({\beta_g\over g}\right),}
where $\beta_g$ is given by the NSVZ formula Eq.~\An, and ${\cal O}$ is as 
defined in Eq.~\Ajb.  It is intriguing that the same operator connects 
$\beta_M$ with $\beta_g$ and $\gamma_1$ with $\gamma$. 
This implies that Yamada's rules for calculating $\gamma_1$ from
$\gamma$ can also be used to derive $\beta_M$ from $\beta_g/g$, which
was left as an open question in Ref.~\yam.
As we shall show explicitly below, Eq.~\Ao\ is in fact also true 
in DRED.

It was shown in 
Ref.~\ref\jjna{I.~Jack, D.R.T.~Jones and C.G.~North, \plb386 (1996) 138\semi
I.~Jack, D.R.T.~Jones and C.G.~North, \npb486 (1997)479}
that $\beta_g^{\NSVZ}$ could be transformed into $\beta_g^{\DRED}$
by a redefinition of the form $g\to g' (g,Y,Y^*)$. 
Let us assume that $\beta_M^{\DRED}$ and $\beta_M^{\NSVZ}$ are related by 
this redefinition, together with $M\to M'(g, h, M, Y,Y^*)$. 
We call the scheme related to DRED by these redefinitions the NSVZ scheme; note 
that $Y$ and $h$ are not redefined.
If $Y$ were to change, for
instance, then the result Eq.~\Aia\ for $\beta_Y$, which follows from the 
non-renormalisation theorem, would not in general be true in both 
schemes\ref\jjn{I.~Jack, 
D.R.T.~Jones and C.G.~North, \npb 473 (1996) 308}. Similarly,
if $h$ were to change, then Eq.~\Ai\ could not be true in both schemes. 
We can now derive a relation between the gaugino mass in the NSVZ
scheme and that in DRED. We will use $g$, $M$, etc to denote the couplings
in the NSVZ scheme, and $g'$, $M'$, etc to denote the couplings in DRED.
The first ingredient is the fact that
$\gamma$ and $\gamma_1$ transform in general according to:  
\eqn\Ap{\eqalign{
\gamma'(g',Y', Y^{\prime *})&=\gamma(g,Y,Y^*),\cr
\gamma_1^{\prime}(g',Y',Y^{\prime *},M',h')&=\gamma_1(g,Y,Y^*,M,h),\cr}}
which follows from the definitions of $\gamma$ and $\gamma_1$. 
Writing
\eqn\Aq{\eqalign{
\gamma'_1(g',Y, Y^*, M',h)&=\left(M'g^{\prime2}{\partial\over{\partial 
g^{\prime2}}}
-h^{lmn}{\partial\over{\partial Y^{lmn}}}\right)\gamma'(g',Y,Y^*)\cr
\gamma_1(g,Y,Y^*,M,h)&=\left(Mg^2{\partial\over{\partial g^2}}-h^{lmn}{\partial
\over{\partial Y^{lmn}}}\right)\gamma(g,Y,Y^*),\cr}}
and using Eq.~\Ap, we find
\eqn\Ar{
gM=g^{\prime}M^{\prime}{{\partial g(g',Y,Y^*)}\over{\partial g'}}
-2h^{ijk}{{\partial g(g',Y,Y^*)}\over{\partial Y^{ijk}}}.}

We can show that the result Eq.~\Ao, which was derived 
in the NSVZ scheme,
is true in DRED, or indeed any scheme related to NSVZ by a redefinition
of $g$ and $M$ alone. After writing
\eqn\Ara{\eqalign{
\beta'_{M'}&={\partial M'\over{\partial M}}\beta_{M}+
{\partial M'\over{\partial g}}\beta_g +
{\partial M'\over{\partial h^{lmn}}}\beta_h^{lmn} +
{\partial M'\over{\partial Y^{lmn}}}\beta_Y^{lmn} +
{\partial M'\over{\partial Y_{lmn}}}\beta_{Ylmn},\cr
\beta'_{g'}&={\partial g'\over{\partial g}}\beta_g+
{\partial g'\over{\partial Y^{lmn}}}\beta_Y^{lmn} +
{\partial g'\over{\partial Y_{lmn}}}\beta_{Ylmn},\cr}}
and using the fact that ${\cal O}$ is form invariant under 
change of scheme (see Eqs.~\Ap, \Aq), 
the proof is largely an exercise in partial differentiation. However,
we need Eq.~\Ai, together with
\eqn\Arb{
M^{ijk}{\partial g'\over{\partial Y^{ijk}}}=
M_{ijk}{\partial g'\over{\partial Y_{ijk}}},}
where $M^{ijk}=\gamma_1^i{}_lY^{ljk}+\gamma_1^j{}_lY^{ilk}
+\gamma_1^k{}_lY^{ijl}$. We
stress once again that $M_{lmn}=(M^{lmn})^*$.
Eq.~\Arb\ follows from the fact that $g'$ consists of a contraction of
equal numbers of $Y$ and $Y^*$, with possible insertions of group matrices
with which $\gamma_1$ commutes.
   
We now give some explicit results up to three loop order.
At one loop we have
\eqn\As{
\lf\beta_g^{(1)}=g^3Q,\qquad
\lf\beta_M^{(1)}=2g^2QM,\qquad
\lf\gamma^{(1)i}{}_j=P^i{}_j,}
where
\eqn\At{
P^i{}_j={1\over2}Y^{ikl}Y_{jkl}-2g^2C(R)^i{}_j,}
and $Q$ was defined above. 
We can now use Eq.~\Ao\ to obtain $\beta_M$ at two-loops, 
using
\eqn\Ax{
 \llf\beta_g^{(2)}=2g^5C(G)Q-2g^3r^{-1}C(R)^i{}_jP^j{}_i .}
We find
\eqn\Ay{
\llf\beta_M^{(2)}=g^2\Bigl(8C(G)QMg^2-4r^{-1}C(R)^i{}_jP^j{}_iM
+2r^{-1}X^i{}_jC(R)^j{}_i\Bigr),}
where $X^i{}_j$ is defined by 
\eqn\Av{
X^i{}_j=h^{ikl}Y_{jkl}+4g^2MC(R)^i{}_j=-2{\cal O}P^i{}_j.}

At two loops we expect the NSVZ and DRED results to agree; the NSVZ and 
DRED results for $\beta_g$ only start to differ at three 
loops\jjna,
and therefore we may write 
\eqn\Az{
g'=g+\delta g, \quad\hbox{and}\quad M'=M+\delta M }
where $\delta g$ starts at two-loop order in $g$. Eq.~\Ar\ then implies that 
$\delta M$ also starts at two loops, and the NSVZ and DRED results for $\beta_M$
should also agree up to two loops. Indeed, Eq.~\Ay\ agrees with earlier DRED
calculations\ref\yama{Y.~Yamada, \plb 316 (1993) 109;
\prl72 (1994) 25}. 

We now turn to the three-loop calculation of $\beta_M$. $\beta_g$ at three
loops is given according to Eq.~\An\ by
\eqn\Ba{\eqalign{
\lllf\beta^{(3)\NSVZ}_g &=
4g^7 Q C(G)^2 -4g^5 C(G) r^{-1} \lf \tr\left[ \gamma^{(1)}C (R)\right]
\cr&-2g^3 r^{-1} \llf \tr\left[ \gamma^{(2)}C (R)\right],\cr}}
where
\eqn\Bb{
\llf\gamma^{(2)i}{}_j=-[Y_{jmn}Y^{mpi}+2g^2C(R)^p{}_j\delta^i{}_n]P^n{}_p+
2g^4C(R)^i{}_jQ.}
Eq.~\Ao\ then leads immediately to
\eqn\Bc{\eqalign{
\lllf\beta_M^{\NSVZ(3)}=&r^{-1}M\Biggl\{4g^2\tr[S_4C(R)]
+16g^4\tr[PC(R)^2]-16g^4C(G)\tr[PC(R)]\cr
&-24g^6Q\tr[C(R)^2]\Biggr\}+24g^6QC(G)^2M\cr
& -r^{-1}\Biggl\{4g^2h^{imp}Y_{jmn}P^n{}_pC(R)^j{}_i+2g^2Y^{imp}Y_{jmn}
X^n{}_pC(R)^j{}_i\cr
& +4g^4\tr[XC(R)^2]-4g^4C(G)\tr[XC(R)]\Biggl\},\cr}}
where $(S_4)^i{}_j=Y^{imp}Y_{jmn}P^n{}_p$.
We have given this result the superscript NSVZ since we know that the NSVZ and 
DRED results will be different at this loop order. 
The $\delta g$ required to transform $\beta_g$ from
NSVZ to DRED is given by\jjna
\eqn\Bd{
\delta g = (\lf)^{-2}{1\over2}g^3
\left[r^{-1}\tr\left[PC(R)\right]-g^2QC(G)\right].}
At the lowest non-trivial order, writing $g'$ and $M'$ according to Eq.~\Az,
Eq.~\Ar\ becomes
\eqn\Be{
\delta M =\left({\partial \delta g\over{\partial g}}-{\delta g\over g}\right) M
-2g^{-1}h^{ijk}{\partial \delta g\over{\partial Y^{ijk}}}.}
Using Eq.~\Bd, we find
\eqn\Bf{\eqalign{
\llf\delta M &=Mg^2\left\{r^{-1}\tr[PC(R)]-2g^2r^{-1}\tr[C(R)^2]
-2g^2QC(G)\right\}\cr
& -g^2(2r)^{-1}h^{ikl}Y_{jkl}C(R)^j{}_i.\cr}}
The consequent change in $\beta_M$ is given by 
\eqn\Bh{\eqalign{
\lllf\delta \beta_M=&r^{-1}M\Biggl\{2g^2\tr[S_4C(R)]+2g^2\tr[P^2C(R)]
+8g^4\tr[PC(R)^2]\cr
&-12g^6Q\tr[C(R)^2]\Biggr\}-6g^6Q^2C(G)M\cr
&-r^{-1}\Biggl\{2g^2h^{imp}Y_{jmn}P^n{}_pC(R)^j{}_i
+g^2Y^{imp}Y_{jmn}X^n{}_pC(R)^j{}_i\cr
& +2g^2\tr[XPC(R)]+2g^4\tr[XC(R)^2]\Biggr\}.\cr}}
We therefore find
\eqn\Bi{\eqalign{
\lllf\beta_M^{\DRED(3)}=&\lllf\beta_M^{\NSVZ(3)}+\lllf\delta \beta_M\cr
=&r^{-1}M\Biggl\{6g^2\tr[S_4C(R)]
+2g^2\tr[P^2C(R)]+24g^4\tr[PC(R)^2]\cr
& -16g^4C(G)\tr[PC(R)]-36g^6Q\tr[C(R)^2]\Biggr\}\cr
& +6g^6M\left\{4QC(G)^2-Q^2C(G)\right\}
 -r^{-1}\Biggl\{6g^2h^{imp}Y_{jmn}P^n{}_pC(R)^j{}_i\cr&+3g^2Y^{imp}Y_{jmn}
X^n{}_pC(R)^j{}_i
 +2g^2\tr[XPC(R)]+6g^4\tr[XC(R)^2]\cr&-4g^4C(G)\tr[XC(R)]\Biggr\}.\cr}}
It is easy to check that this result may be derived using Eq.~\Ao\ starting
from $\beta_g^{\DRED(3)}$ as given in Ref.~\jjna, in accord with our general 
results. Moreover, we have checked several terms in Eq.~\Bi\ by explicit
calculation; we hope to report on a full computation in due course. 

It is readily verified that Eqs.~\Ai, \Aj, \Ajb, \As\ and \Bb\ lead to the
one and two-loop results for $\beta_h$ and $\beta_b$ quoted in 
Refs.~\ref\mv{S.P.~Martin and M.T.~Vaughn, \prd50 (1994) 2282}, \yam\ and \jj.
Moreover, we also find that Eq.~\Ajc\ leads to the correct one-loop 
$\beta$-function for $m^2$,
\eqn\Awa{\eqalign{
\lf[\beta_{m^2}^{(1)}]^i{}_j=&{1\over2}Y_{jpq}Y^{pqn}(m^2)^i{}_n
+{1\over2}Y^{ipq}Y_{pqn}(m^2)^n{}_j
+2Y_{jpq}Y^{ipr}(m^2)^q{}_r\cr 
&+h_{jpq}h^{ipq}-8g^2MM^*C(R)^i{}_j.\cr}}
However, the two-loop case requires more discussion. DRED leads to 
the introduction of $\epsilon$-scalars which maintain the equality of bosonic
and fermionic degrees of freedom. In the softly-broken theory, 
the $\epsilon$-scalars acquire a mass
under renormalisation,which  means that an $\epsilon$-scalar mass parameter 
should be introduced,  which
impacts on the calculation of $\beta_{m^2}$. It has been shown\jjmvy\ 
that there is a scheme in which the dependence on the $\epsilon$-scalar mass 
decouples from $\beta_{m^2}$. 
For this scheme, using the two-loop results from 
Ref.~\jj, we find
\eqn\Awd{\eqalign{
[\beta_{m^2}^{(2)}]^i{}_j=&\left[2{\cal O}\bar {\cal O}+2M\bM g^2{\partial
\over{\partial g^2}} +\tY_{lmn}{\partial\over{\partial Y_{lmn}}}
+\tY^{lmn}{\partial\over{\partial Y^{lmn}}}\right]\gamma^{(2)i}{}_j\cr
& +8g^4(\lf)^{-2}SC(R)^i{}_j,\cr}}
where $S$ is defined by
\eqn\Awc{
S\delta_{AB}=(m^2)^k{}_l(R_AR_B)^l{}_k
-MM^* C(G)\delta_{AB},}
and arises because there are divergent one-loop contributions to 
the $\epsilon$-scalar mass. 
This shows, as mentioned earlier, that Yamada's rules do not take into account
the $\epsilon$-scalar mass. However, one might hope that there exists  
another renormalisation scheme, without continuation away 
from four dimensions, such that Yamada's rules would apply exactly. 
There does not exist, however,  an appropriate 
redefinition which simply transforms away the term in $S$ in Eq.~\Awd.
Consequently the precise treatment of $m^2$ within the NSVZ scheme remains 
unresolved.  

We turn now to the issue of finite $N=1$ theories. 
It is believed that a supersymmetric theory which is finite at one-loop order
is finite to all orders. It has been known for some time that if a 
supersymmetric 
theory is finite to $L$ loops (i.e. $\beta_g$ and $\gamma$ both vanish to this
order) then $\beta_g$ will also vanish at $L+1$ loops\ref\PW{A.J.~ Parkes 
and P.~West, \npb256 (1985) 340}
\ref\gmz{M.T.~Grisaru, B.~Milewski and D.~Zanon, \plb 155 (1985)357}. 
Moreover there are arguments\ref\lpsk{C.~Lucchesi, O.~Piguet and 
K.~Sibold, \plb 201 (1988) 241\semi
C. Lucchesi, hep-th/9510078\semi
A.V.~Ermushev, D.I.~Kazakov and O.V.~Tarasov, \npb 281 (1987) 72}\
that in a one-loop finite theory, $\gamma$ can be
transformed to zero to all orders (though it is unclear to what extent 
the argument
depends on the number of fields relative to the number of independent 
couplings\ref\tjnc{D.R.T.~Jones, \npb 277 (1986) 153}). 
It has been verified explicitly\jjn\ that $\gamma$ can indeed 
be transformed
to zero up to three loops in a one-loop finite theory. It is interesting to ask
whether similar results hold in the soft-breaking sector. 
It has been known for some time\ref\jmy{D.R.T.~Jones,
L.~Mezincescu and Y.-P.~Yao, \plb148 (1984) 317.}\ that in a softly-broken
supersymmetric theory for which 
$P^i{}_j=Q=0$, thus rendering the supersymmetric part of the theory one-loop 
finite, the relations 
\eqn\Ca{\eqalign{
h^{ijk}=&-MY^{ijk},\cr
b^{ij}=&-{2\over3}M\mu^{ij}\cr}}
suffice to render $\beta_h^{ijk}$ and $\beta_b^{ij}$ zero at one loop. It was 
later shown\jj\ that in 
fact $\beta_h^{ijk(2)}$ and $\beta_b^{ij(2)}$ then also
vanish. It is easy to see how this works. On imposing 
Eq.~\Ca, we have
\eqn\Cb{
{\cal O}=M\left(g^2{\partial\over{\partial g^2}}+Y^{lmn}{\partial
\over{\partial Y^{lmn}}}\right).}
From Eqs.~\Aj, \Ao, it then follows that (at $L$ loops)
\eqn\Cc{
\gamma_1^{(L)}=LM\gamma^{(L)}, \qquad \beta_M^{(L)}=2L\beta_g^{(L)}M/g.}
Thus, since $P^i{}_j=Q=0$ implies $\gamma^{(2)}=\beta_g^{(2)}=0$,
we have immediately that one-loop finiteness implies two-loop finiteness in
the softly-broken case. However, unfortunately it is not clear how to extend
this argument to higher orders. 
The scheme for which $\gamma^{(3)}$ vanishes in the
one-loop finite case is related to DRED by a redefinition of $Y$\jjn, and so in
this scheme Eq.~\Aj\ no longer holds.
 
On the other hand, we can directly demonstrate that one-loop
finiteness implies the all-orders vanishing of $\beta_g$ and $\beta_M$,
irrespective of whether $\gamma$ or $\gamma_1$ vanish. The redefinition
\eqn\Cd{
g'=g+{1\over6}{g\over\lf}r^{-1}\tr P }
implies, upon differentiating with respect to $\mu$, and using 
Eqs.~\An\ and \At, that 
\eqn\Cg{
\lf\beta_g'(g',Y)=Qg^2\left(g-{2\over3}\beta_g\right)+r^{-1}
\left(g\tr[\gamma P]+{1\over6}\beta_g\tr P\right).}
It is clear from this expression that $\beta'_g(g',Y')$ 
vanishes to all orders
in the one-loop finite case, where $P^i{}_j=Q=0$.

Similarly, the redefinition 
\eqn\Ch{
M'=M-{1\over6}{1\over\lf}r^{-1}\tr X }
leads, using Eqs.~\Ai,\Aia,  \At, \Am\ and \An, to
\eqn\Ck{\eqalign{
\lf{\beta}'_M(M',h,Y,g)=&2g^2MQ\left(1-{2\over3}{{\beta}_g\over g}\right)
-{2\over3}g^2Q{\beta}_M\cr
& -r^{-1}\tr[\gamma X]+2r^{-1}\tr[\gamma_1 P].\cr}}
Again it is manifest that for a one-loop finite theory, for which $P^i{}_j
=X^i{}_j=Q=0$, ${\beta}'_{M'}=0$ to all orders.  
It is a simple matter
to verify that Eqs.~\Cd\ and \Ch\ are consistent with Eq.~\Be.

In conclusion: we have reformulated the HS result into an elegant relation 
between $\beta_M$ and $\beta_g$ which is true in both the NSVZ scheme 
and (remarkably) DRED. We have also shown that in a one-loop finite 
theory there exists a redefinition of $g$ such that $\beta_g = \beta_M=0$ 
to all orders. Issues related to the soft scalar mass $m^2$ and the 
redefinition of $Y$ needed to make $\gamma^{(3)}$ vanish in a one-loop 
finite theory remain to be resolved. 
\listrefs
\bye